\documentclass[11pt, a4paper]{article}

\usepackage{jheppub}

\usepackage{mathrsfs}
\usepackage[T1]{fontenc}
\usepackage{mathpazo}
\usepackage{setspace}
\usepackage{amsfonts}
\usepackage{amssymb}
\usepackage{amsmath}
\usepackage{epsfig}
\usepackage{latexsym}
\usepackage{color}
\usepackage{graphicx}
\usepackage{nicefrac}
\usepackage[latin1]{inputenc}
\usepackage{pstricks}
\usepackage{slashed}
\usepackage{multirow}

\usepackage{hyperref}

\def\be{\begin{equation}}
\def\ee{\end{equation}}
\def\ba{\begin{eqnarray}}
\def\ea{\end{eqnarray}}

\def\k{\kappa}

\def\a{\alpha}

\def\b{\beta}

\def\g{\gamma}
\def\G{\Gamma}
\def\d{\delta}

\def\e{\epsilon}

\def\C{\Chi}

\def\om{\omega}

\def\L{\Lambda}
\def\s{\sigma}

\def\IR{\relax{\rm I\kern-.18em R}}

\def\inv{^{\raise.0ex\hbox{${\scriptscriptstyle -}$}\kern-.05em 1}}

\title{Duality Invariance: From M-theory to Double Field Theory}

\author[a]{Daniel C. Thompson}

\affiliation[a]{Theoretische Natuurkunde, Vrije Universiteit Brussel, and 
The International Solvay Institutes\\
Pleinlaan 2, B-1050, Brussels, Belgium}

\emailAdd{dthompson@tena4.vub.ac.be}

\abstract{We show how the duality invariant approach to M-theory formulated by Berman and Perry relates to the double field theory proposed by Hull and Zwiebach. In doing so we provide suggestions as to how Ramond fields can be incorporated into the double field theory.  We find that the standard
dimensional reduction procedure has a duality invariant (doubled) analogue in which the gauge fields of the doubled Kaluza-Klein ansatz  encode the Ramond potentials.   We identify the internal gauge index of these gauge fields with a spinorial index of $O(d,d)$.}

\begin{document}

\maketitle
\flushbottom

 \def\H{ {\cal H}}
   \def\G{ {\cal G}}
    \def\C{ {\cal C}}
    \def\M{{\cal M}}
    \def\pd{\partial}

\section{Introduction}

The low energy limit of M-theory is eleven-dimensional supergravity, which upon dimensional reduction on a circle yields type IIA ten-dimensional supergravity (which is in turn the low energy limit of the type IIA superstring) \cite{Witten:1995ex}.  Reduction of the eleven-dimension supergravity theory  \cite{Cremmer:1978km}  on higher tori result in lower dimensional supergravity theories which possess an extremely rich structure of  hidden-symmetries \cite{CremmerLowerD}.  For instance,   the result of reduction on a $T^4$ is a seven-dimensional theory with an $SL(5)$ symmetry,  for a $T^{5}$ reduction the duality group is $SO(5,5)$ and proceeding further one finds hidden symmetries classified by the exceptional Lie-algebras (and eventually their affine extensions).  Devolping a complete understanding of such hidden symmetries and indeed the extent to which these symmetries play an underlying role of the full uncompactified theory has long been a topic of study \cite{deWitHidden,West:2001as}.   For many years it has been suggested that these symmetries might be made more manifest by the inclusion of extra dimensions (above the eleven) to account for the central charges of the supersymmetry algebra, see for instance \cite{deWit:2000wu}. 

Some progress in this direction has been made recently \cite{ Hillmann:2009zf, Berman:2010is,Berman:2011pe}.  In \cite{Berman:2010is}  a canonical quantisation approach along the lines of \cite{Arnowitt:1962hi,DeWitt:1967yk}  was used to show that the \emph{uncompactified} theory can be recast in a way that makes the hidden symmetry group manifest.   To be more precise this is achieved for a theory of gravity together with a three-form potential defined in $d=1+4$ dimensions replicating the bosonic sector of eleven-dimensional supergravity.  Drawing inspiration from the structures seen in generalised geometry \cite{Gualtieri:2003dx,Hitchin:2004ut,Hull:2007zu,Pacheco:2008ps},  the Hamiltonian of this theory was rewritten in such away that the $SL(5)$ hidden symmetry group is made manifest.   To achieve this the four spatial coordinates were augmented with an additional six extra coordinates (which can be though of as being due to membrane charges).     This result was later extended to make the $SO(5,5)$ symmetry of the $d=1+5$ dimensional  version manifest  \cite{Berman:2011pe}; in this case an additional eleven spatial coordinates need to be included (which arise from the ten membrane charges and a single five-brane charge).     

In parallel to this has been the recent renewed interest and development of a duality invariant target space formulation of string theory known as double  field theory (DFT) \cite{DFT}  building on important preceding works by Tseytlin \cite{TseytlinDoubled} and  Siegel  \cite{Siegel:1993th,Siegel:1993xq}. In this approach a particular truncation of bosonic closed string field theory on a torus $T^n$ was shown to give rise to an background independent  effective action exhibiting manifest T-duality invariance.  This effective action has double the usual number of coordinates i.e.  the $n$ regular coordinates $x$ are supplemented by their T-dual partners $\tilde{x}$  (which can be though of as arising from string winding modes) and the fields can depend on both $x$ and $\tilde{x}$.    

An alternative approach is  based on a proposal made in 2003, \cite{West:2003fc},  to consider the non-linear realisation of $E_{11}$ and its first fundamental representation.  This introduced an extension of space-time to include extra coordinates in one-to-one correspondence with brane charges and a generalised veilbein \cite{West:2003fc, Kleinschmidt:2003jf, West:2004kb, West:2004iz, Cook:2008bi}.    In \cite{West:2010ev}, West constructed this non-linear realisation and, at level zero from the IIA point of view, the dynamics contained the NS fields and agreed with that of the double field theory.  This work was extended to level one and the RR sector of IIA supergravity in \cite{Rocen:2010bk}.   

A precursor to  DFT is the development of a world sheet duality invariant theory which exhibits many of the structures found in the double field theory - a selection of relevant work on this can be found in \cite{Duff:1989tf,Duff:1990hn,TseytlinDoubled,TseytlinDoubled2, HullDoubled}, (see \cite{Thompson:2010sr} for a review and further references).   Recently, it has been shown how, at least in a certain simplified scenario, the  DFT equations of motion which take the form of the vanishing of some generalised Ricci tensor arise directly as constraints of conformal invariance in these world sheet approaches \cite{BFE}.  

In this paper we  examine how the duality invariant M-theory can be related directly to double field theory via dimensional reduction.   Of course, that each is equivalent to its standard formulation means  there surely is a some relation  however it is by no means evident exactly how the mechanics of this should work.  A further reason for this work, and indeed the original motivation, is to provide further perspective on the inclusion of Ramond-Ramond p-forms into DFT.   The fact that the M-theory treats fields that descend to the RR sector democratically with those that descend to NS fields underlies this thinking. Certainly the limitations of current string field theory technology  render a direct approach to this problem unviable. 

For the reader's  convenience  we now give a short summary of the key results within.  Firstly in the would-be NS sector we shall see how the standard KK ansatz (together with appropriate  Weyl rescaling) for the M-theory fields naturally gives rise to a doubled Kaluza-Klein ansatz in the doubled geometry.   The external metric of this ansatz is simply the $O(d,d)$ generalised metric $\H$ of the DFT. Unlike the standard dimensional reduction the internal metric $\G$  of the doubled KK ansatz still depends on the geometric data (the IIA metric $h$ and two-form $b$) however in a different representation; the internal metric acts on spinors of  $O(d,d)$.    Restricting our attention to the NS sector we find that the dimensional reduction gives rise to the Lagrangian of DFT.  This is only possible  through some rather non trivial cancelations and relationships between $\G$ and $\H$.     

Secondly, when we include the RR sector we find  that in the doubled geometry the RR fields elegantly assemble themselves into a single KK `gauge' field  of the Doubled KK ansatz.    These gauge fields carry a $O(d,d)$ vector index and an internal $Spin(d,d)$ spinorial index.   Their kinetic terms are relatively simple and suggestive of potential generalisation.  The coupling of the gauge field spinorial indices to the NS sector is by means of the spinorial representation of the generalised metric $\G$.       
 
The idea that spinorial representations of $O(d,d)$ are relevant to RR fields has some precedent in the literature.  For instance using non-linear realisations of exceptional symmetry algebra \cite{West:2003fc, West:2010ev, Rocen:2010bk} it is shown that RR fields can be packaged as spinors of $O(d,d)$ and for the  case of $O(9, 9)$ in  \cite{Kleinschmidt:2004dy}.  This idea can also be seen in  \cite{Fukuma:1999jt} wherein the RR fields couple to the NS sector by means of the same spinorial representation of the generalised metric $\G$.

There are several subtleties that require careful consideration and mean that the relationship between the doubled M-theory and DFT is less evident than one might immediately assume. A principal complication is that time is treated quite differently in the two approaches; in the M-theory time is singled out and treated distinctly from spatial coordinates which are doubled whereas the DFT not only maintains covariance but actually doubles the time direction implicitly.  A further  complication arises because we are not actually working with eleven-dimensional supergravity but rather a lower dimensional cousin.  Since the dilaton $\phi$ produced by KK scalar in the dimensional reduction is sensitive to this dimensionality we find that one can reproduce DFT only up to an additional quadratic dilaton term (nonetheless the terms involving the DFT doubled dilaton $d$ do come out correctly).

The structure of the remainder of this paper is as follows: In the next section we review the standard dimensional reduction procedure paying careful attention to how dimensionality alters things. In section 3, we turn our attention to the $SL(5)$ invariant M-theory. We will first briefly review the construction of    \cite{Berman:2010is} and then consider its dimensional reduction. After comparing the NS sector to that of DFT we turn to the RR sector.   In section 4 we repeat this analysis for the richer case of the $SO(5,5)$ invariant M-theory.  We close in section 5 with a short discussion.   
\\
\\
\emph{ A brief note on notation: Hatted indicies correspond to M-theory indices and unhatted to string theory.   Uppercase indices correspond to doubled/generalised indices and lowercase to the standard formulation.}

\section{Dimensional reduction away from criticality} 
Although the relation between eleven-dimensional supergravity and type IIA supergravity in ten dimensions is well known \cite{Witten:1995ex} in this paper we shall need to work away from criticality and so  revisit the dimensional reduction procedure for arbitrary dimension.  As we shall see, the dilaton normalisation and kinetic term are dimensionally sensitive, something which will impact the discussion later. 

We start in $n+1$ dimensions with the action
\be
S^{(n+1)} = \frac{1}{\k_{(n+1)}^2} \int d^{n+1}x \sqrt{G} \left( R[G] - \frac{1}{48} F_{\hat{n}\hat{m} \hat{p}\hat{q} } F^{\hat{n}\hat{m} \hat{p}\hat{q}  }  \right) \ , 
\ee
and we dimensionally reduce on an $S^{1}$ labelled by coordinate $z$  with the ansatz 
\be
G_{\hat{m} \hat{n} } =   \left( \begin{array}{c    c}  \tilde{g}_{ij} & 0 \\ 0 & e^{2 \g} 
  \end{array} \right) \ ,  \quad  C_{ i j z} =  b_{i j}  \ , 
\ee
to find 
\be
S^{(n)} = \frac{1}{\k_{(n )}^2} \int d^{n}x \sqrt{\tilde{g}} e^{\g} \left( R[\tilde{g}]   - 2 (\pd \g)^2 - 2 \tilde{\nabla}^2 \g - \frac{1}{12} H_{ijk}H^{ijk}e^{2 \g} \right) \ . 
\ee
In order that all the terms exhibit homogeneous scaling we perform a Weyl rescaling $\tilde{g}_{ ij} = e^{ - \g} g_{ij}$.  Making use of the identities \eqref{Weylresc} the action is then given by
\be
S^{(n)} = \frac{1}{\k_{(n )}^2} \int d^{n }x \sqrt{g} e^{\left(2 - \frac{n}{2} \right)\g} \left( R[g]   + \left(n - 4 -\frac{1}{4}(n-1)(n-2)  \right)(\pd \g)^2  + (n - 3) \nabla^2 \g - \frac{1}{12} H_{ijk}H^{ijk} \right) \  . 
\ee 
At this stage we may integrate by parts to find
\be
\label{sred1}
S^{(n)} = \frac{1}{\k_{(n )}^2} \int d^{n }x \sqrt{g}  e^{\left(2 - \frac{n}{2} \right)\g}  \left( R[g]  + \frac{1}{4}(n-6)(n-1) \pd_i \g \pd^i \g  - \frac{1}{12} H_{ijk}H^{ijk} \right) \   . 
\ee
We now perform a field redefinition to identify the dilaton with the KK scalar according to 
\be
\label{gammadilaton}
e^{\left(2 - \frac{n}{2} \right)\g} = e^{-2 \phi} \ ,
\ee
so that the action takes its final form 
\be
\label{sred}
S^{(n)} = \frac{1}{\k_{(n )}^2} \int d^{n }x \sqrt{g}  e^{ - 2 \phi }  \left( R[g]  +  N[n] \pd_i \phi \pd^i \phi  - \frac{1}{12} H_{ijk}H^{ijk} \right) \ ,
\ee
with the coefficient
\be
N[n] = (n- 6)(n-1) \left(2 - \frac{n}{2} \right)^{-2} \ .
\ee

Of interest in this paper are the cases $n=10$ corresponding to actual eleven-dimensional supergravity, $n=4$ corresponding to the $SL(5)$ duality  invariant theory of \cite{Berman:2010is} and $n=5$ corresponding to the $O(5, 5)$ duality invariant theory in \cite{Berman:2011pe}.

For the $n=10$ case we recover the standard famous relation that 
\be
n=10 \Rightarrow \g = \frac{2}{3} \phi 
\ee
and the normalisation $N[10] = 4$ is such that the dimensionally reduced \eqref{sred} action corresponds to the bosonic NS sector of Type II supergravity.  For the case $n=5$  we have a different relation 
\be
n= 5  \Rightarrow \g = 4 \phi 
\ee
and a normalisation of $N[5]= -16$.  Finally for the case of $n=4$ we encounter some pathology; the prefactor in  \eqref{sred1} vanishes automatically and $N[4]$ correspondingly diverges.  Whilst the action   \eqref{sred1} is still perfectly valid the field redefinition no longer makes sense and if we insist on having an action of the form \eqref{sred} we must enforce:
\be
n= 4  \Rightarrow \g = \phi = 0 \ . 
\ee
One can see that whilst the metric and two-form sectors of the theory are dimensionally insensitive, the dimensional reduction the dilaton sector must be treated with some care when working away from criticality.  Furthermore, since in arbitrary dimensions the DFT is equivalent to \eqref{sred} with the coefficient fixed to be $N[n]=4$ regardless of dimension one sees that there will necessarily be a discrepancy between the dimensionally reduced double M-theory and DFT.

\section{From $SL(5)$ invariant M-theory to $O(3,3)$ DFT} 
\subsection{Duality invariant M-theory} 
The approach of Berman and Perry begins with the action\footnote{Since we shall not be working in eleven dimensions here, there is no Chern Simons term.  In fact, it turns out that a Chern-Simons piece may be included by making a  canonical transformation and the form of the Hamiltonian is essential unaltered.} 
\be
S^{(n+1)} =  \int d^{n+1}x \sqrt{G} \left( R[G] - \frac{1}{48} F_{\hat{m}\hat{n} \hat{p} \hat{q} } F^{\hat{m}\hat{n} \hat{p} \hat{q} }  \right) \ .  
\ee
In the canonical treatment one performs a time slicing ( making the assumption that spacetime can be foliated by equal time sufaces).   The canonical variables of the metric are specified by the lapse function $\a$, shift vector $\b_{\hat{i}}$  and the positive definite spatial metric $G_{\hat{i}\hat{j}}$ and their conjugate momenta $\pi_\a , \ \pi^{\hat{i}}_\b$ and $ \pi^{\hat{i}\hat{j}}$ respectively .    The three form with all legs spatial gives rise to canonical variables $C_{\hat{i}\hat{j} \hat{k}}$ with conjugate momenta $\pi^{\hat{i}\hat{j} \hat{k}}$ and with one temporal leg a further set of canonical variables $C_{\hat{t}  \hat{ i} \hat{j} } = B_{ \hat{ i} \hat{j} }$ with momenta $\pi_B^{ \hat{ i} \hat{j} }$.  The dynamics is that of a constrained system; there are a number of first class constraints that allow the gauge fixing choice of synchronous gauge, i.e.  $\a$, $\b^{\hat{i}}$ and $  B_{ \hat{ i} \hat{j} }=0$ are set to zero.  The complete Hamiltonian, given by the integral of the Hamiltonian constrain, weakly vanishes and is given by
\be
\label{Ham}
H = \int d^n x  G^{-\frac{1}{2}} \left( \pi^{ \hat{ i} \hat{j} } \pi_{ \hat{ i} \hat{j} } - \frac{1}{n-1} \pi^2 +3 \pi^{\hat{i}\hat{j} \hat{k}}\pi_{\hat{i}\hat{j} \hat{k}}  - G\left(  R[G]  -\frac{1}{48} F_{\hat{i}\hat{j} \hat{k}\hat{l} }F^{\hat{i}\hat{j} \hat{k}\hat{l} } \right)  \right) \ ,
\ee      
in which $R[G]$ is the Ricci scalar of the spatial metric and $F$ is the spatial exterior derivative of $C$.  As ever, the time evolution of fields are given by their Poisson bracket with the Hamiltonian.  

The first indications of some novel unexpected structure comes on studying the constraint algebra.  Alongside the Hamiltonian constraint there is a diffeomorphism constraint $\chi_{\hat{i}}$ and a gauge constraint $\chi^{\hat{i}\hat{j} }$.  The first surprise is that the action the diffeomorphism on the three-form is not simply the Lie derivative but rather a combination of Lie derivative and gauge transformation.   A second surprise is that the algebra of diffeomorphisms does not close simply only onto diffeomorphisms but rather up to a strange  field dependent term involving the field strength and gauge constraint:
\be
\{ \chi_{\hat{i}}(x) , \chi_{\hat{j}}(x^\prime) \}_{P.B.} =  \left(\chi_{(\hat{i}} D_{\hat{j})} + \frac{1}{\sqrt{2}}F_{\hat{i}\hat{j} \hat{k}\hat{l} }\chi^{\hat{k}\hat{l}} \right)  \delta(x, x^\prime) \ . 
\ee

The jumping-off point, and a key insight in \cite{Berman:2010is}, is that the rather exotic constraint algebra can be naturally recast in the language of  genralised geometry.  By considering the formal sum of vector fields  $X$ and (in this case) two-forms $\xi$  one enlarges tangent space to $TM \oplus \Lambda^2 T^\ast M$  which may be equiped with the structure of a Lie-algebroid by means of the Courant bracket
\be
[X+\xi , Y+\eta]_C = [X, Y] + L_X \xi -  L_Y \eta - d(\iota_X \eta - \iota_Y \xi ) \ ,
\ee
in which, as usual, $L$ represents the Lie derivative,  $\iota$ is the interior product and $d$ the exterior derivative.   This structure can be further modified to include a twisting
\be
[X + \xi , Y+ \eta ]_T = [X+\xi , Y+\eta]_C  + \frac{1}{\sqrt{2} } \iota_Y \iota_X F  \ , 
 \ee
where $F = d  C$ .   Remarkably the constraint algebra of diffeomorphisms and gauge transformations has exactly this form.    This,  together with the much earlier worldsheet study of Duff and Lu \cite{Duff:1990hn} suggested that one might be able to harness the power of generalised geometry to recast the canonical theory \eqref{Ham} in an insightful way.

In the case at hand  the generalised geometry contains ten coordinates $X^{\hat{M}}$ (i.e. the four standard ones $x^{\hat{a}}$ and six others $y_{\hat{a}\hat{b}}$ representing dual two-cycles) and their derivatives $\pd_{\hat{M}} = (\pd_{\hat{a}}, \pd^{\hat{a} \hat{b}} )$.   The generalised metric for this  this ten dimensional extended space is given by
\be
\label{genmetric}
M_{\hat{M} \hat{N}}= \left( \begin{array}{cc}
   G_{\hat{a}\hat{ b}} + \frac{1}{2} C_{\hat{a} \hat{e}\hat{f}} C_{\hat{b}}{}^{\hat{e} \hat{f}}  & \frac{1}{\sqrt{2}} C_{\hat{a}}{}^{\hat{k}\hat{l} } \\ 
 \frac{1}{\sqrt{2}} C_{\hat{b}}{}^{\hat{m}\hat{m}}&  \frac{1}{2}(G^{\hat{m}  \hat{k}} G^{\hat{n}\hat{l}}  -G^{\hat{m}  \hat{l}} G^{\hat{k}\hat{n}}    )  
  \end{array} \right) \ . 
\ee
Since the metric parametrises the symmetric coset $SL(5)/SO(5)$ a reformulation of  \eqref{Ham} in terms of this object makes manifest the global $SL(5)$ hidden symmetry of M-theory.

The dynamics for this  generalised metric are given by a Hamiltonian
\be
H_{BP} = T_{BP} + V_{BP}
\ee
with the  kinetic terms 
\be
\label{TBP}
T_{BP} = -\sqrt{G} \left(  \frac{1}{12} tr (\dot M^{-1} \dot M ) + \frac{1}{12} (tr (  M^{-1} \dot M ))^2 \right)
\ee   
and a potential  
\be
\label{VBP}
\frac{1}{\sqrt{\det{G}} } V_{BP} = V_1  + V_2 + V_3 + V_4 
\ee
with 
\ba
 V_1 =  \frac{1}{12} M^{\hat{M}\hat{N}} (\partial_{\hat{M}} M^{\hat{K}\hat{L}}) (\partial_{\hat{N}} M_{\hat{K}\hat{L}}) \ ,  &\quad& 
V_2 = - \frac{1}{ 2} M^{\hat{M}\hat{N}} (\partial_{\hat{N}} M^{\hat{K}\hat{L}}) (\partial_{\hat{L}} M_{\hat{K}\hat{M}}) \ ,   \\
 V_3 =  \frac{1}{12} M^{\hat{M}\hat{N}} ( M^{\hat{K}\hat{L}} \partial_{\hat{M}} M_{\hat{K}\hat{L}})( M^{\hat{R}\hat{S}} \partial_{\hat{N}} M_{\hat{R}\hat{S}})\ , &\quad&
V_4 =  \frac{1}{4} M^{\hat{M}\hat{N}}  M^{\hat{P}\hat{Q}}  ( M^{\hat{K}\hat{L}} \partial_{\hat{P}} M_{\hat{K}\hat{L}})(   \partial_{\hat{M}} M_{\hat{N}\hat{Q}})\ . \nonumber
\ea
Upon unpacking this expression rather carefully and invoking the `section condition' that $\pd^{ab} = 0$ one can show that this Hamiltonian is equivalent (upto surface terms) with the original one in \eqref{Ham}.   Whereas the canonical Hamiltonian \eqref{Ham} has a clear geometric understanding (the potential contains a Ricci Scalar) it is, at present, unclear how best to interpret this potential -- it is certainly not the standard notion of Ricci tensor.     In this derivation no isometry properties or compactness properties were needed -- the symmetry is intrinsic to the \emph{uncompactified} theory.

\subsection{The doubled Kaluza--Klein ansatz}
Now let us consider how the dimensional reduction of section 2 might be applied to the generalised M-theory.  We recall the KK ansatz for an $S_{z}^{1}$ reduction of the $d=1+4$ M-theory.  For the metric we employ
\be
\label{kk1}
G_{\hat{m}\hat{n} } =  \left( \begin{array}{cc}
    e^{-\gamma} h_{m n} + e^{2 \gamma} A_m A_n & A_m e^{2 \gamma}\\ 
  A_n e^{2 \gamma}    &  e^{2 \gamma} 
  \end{array} \right) \ ,
\ee
and for the three-form\footnote{This definition of the three form serves to simplify many of the expressions that follow.} 
\be
\label{kk2}
C^{(3)}_{m n  z} = b_{m n}\ , \quad C^{(3)}_{m n p}  = K_{m n p}  + A_{m} b_{n p} + A_{n} b_{p m}+A_{p} b_{m n} \ .
\ee
Of course, following the discussion in section 2, when we ultimately compare to the DFT we will need to set $\g$ to zero however it is rather illuminating not to do so just yet.   Note that here we are performing the Weyl rescaling and dimensional reduction in one step for reasons which will become immediately apparent. 

The first thing is to understand how the KK ansatz  \eqref{kk1} works on the generalised metric \eqref{genmetric}.  After some tedious manipulation\footnote{For some of the more onerous manipulations we found the symbolic algebra system {\tt Cadabra} \cite{Peeters:2007wn,Peeters:2006kp} to be a helpful tool to verify calculations.} one can  ascertain that the component expressions:
\begin{itemize}
\item For the $4\times 4$ top left block of \eqref{genmetric} :
\ba 
M_{m n} &=&   e^{-\g} {\bf (  h - b h^{-1}b )_{m n} }  + \frac{1}{2 }  e^{ 2\gamma} \left(2 {A}_{m} {A}_{n} \Lambda   + {K}_{m p q} {K}_{n }{}^{p q}  +   {A}_{n} {K}_{m}    +   {A}_{m} {K}_{n }    \right)   \ , \quad \nonumber \\ 
M_{z z} &=& e^{2 \gamma} \Lambda \ ,   \qquad M_{z n} = e^{ 2\gamma} \left( A_n \Lambda  +  \frac{1}{2} K_{n  }  \right) \ , 
\ea
in which $\L = ( 1 + \frac{1}{2} b_{mn } b^{mn}) $ and $K_{n} = K_{n p q} b^{p q}$.
\item For the $4\times 6$ off-diagonal blocks:
\ba
&& M_{z}{}^{ z n}  =   -  \frac{1}{\sqrt{2}} e^{2 \gamma} A^m b_m{}^n \ , \quad M_{z}{}^{m n} =\frac{1}{\sqrt{2}} e^{2 \gamma} b^{m n} \ , \nonumber \\ 
&& M_{m}{}^{z n} =   -\frac{1}{\sqrt{2}}  e^{-\g} {\bf (   bh^{-1} )_m{}^n }  - \frac{1}{\sqrt{2}} e^{2 \gamma}( A_m A^p b_p{}^n +  K_{m p}{}^nA^p  ) \ , \nonumber \\ 
&& M_{m}{}^{n p} =  \frac{1}{\sqrt{2}}  e^{2 \gamma} \left( K_{m}{}^{n p} + A_m b^{n p} \right) \ .
\ea 
\item For the $6\times 6$ bottom right  block:
\ba
&& M^{m n , k l}= \frac{1}{2} e^{2 \gamma} \left(h^{m k}h^{l n} - h^{k n} h^{m l} \right) \ , \nonumber \\
&& M^{z n , z l} = \frac{1}{2} e^{- \gamma}  {\bf h^{n l} }  + \frac{1}{2} e^{2 \gamma} \left(A^2  h^{nl} - A^l A^n \right)\ , \nonumber  \\
&& M^{z n , k l } =  \frac{1}{2} e^{2 \gamma} \left(A^l  h^{k n} - A^k h^{l n} \right) \ . 
\ea
\end{itemize}
Already one can see a certain structure emerge -- the terms are naturally split into two sorts: those scaling with $e^{-\gamma}$ and those scaling as  $e^{+2 \gamma}$. The terms that scale with $e^{-\gamma}$ (highlighted in bold in the above) do not depend on the Ramond fields and appear in the combinations found in the $O(d,d)$ coset metric of the DFT. 

Now we wish to split the $4+6$ dimensional space to a $3+ 3$ dimensional space. We must therefore identify the correct internal space.  For the $x^{\hat{a}}$ coordinates it is obvious that $x^{\hat{a}} = ( x^a , z )$ and we should impose $\frac{\partial}{\partial z} = 0$.  For the $y_{\hat{a} \hat{b} } = ( y_{z a} , y_{a b} )$ it is less clear which of these two sets of three should be considered the internal coordinate. The form of $M_{m}{}^{z n}$ tells us that  $\tilde{x}_n= y_{z n}$ are the three coordinates that we should keep after reduction and that $y_{ m n }$ are internal and we should impose $\frac{\partial}{\partial y_{mn} } = 0$.    Hence the coordinates that survive the dimensional reduction are given by $X^{{M}}= (x^{m}, \tilde{x}_{m} ) = (x^{m} , y_{z m})$. 

The M-theory "section condition" of Berman Perry that $\frac{\partial}{\partial y_{\hat{m}\hat{n} } } = 0$ reduces to saying that the reduced fields do not depend on the $\tilde{x}$ coordinates -- this is exactly in accord with the "strong constraint" of Double  Field Theory.\footnote{A complete understanding of the possible general section conditions for the M-theory is somewhat lacking at the moment.  It would of course be nice to make a link between such a general section condition and the general strong constraint rather than just this one particular solution of it as we have here. }  

In line with the above discussion it suits us to reorder coordinates such that $X^{\hat{M}} = (x^{a}, \tilde{x}_{a},  z, y_{a b}   ) $ i.e. we shuffle components of the M-theory generalised metric about in the following way:
\be
\label{Mgenrelabel} 
M_{\hat{M}\hat{N}} =    \left( \begin{array}{c |c  c}
 {\cal M}_{M N} & {\cal A}_{M} & {\cal B}_{M}^{r s}  \\ \hline
{\cal A}_{N}  &  M_{zz} & M_{z}{}^{r s} \\
 {\cal B}_{N}^{p q} &  M^{pq}{}_{z} & M^{p q , r s}  
  \end{array} \right) \ , 
\ee
where the top left block now represents the reduced generalised metric in `external' directions and the bottom right represents the `internal' metric. For the reduced generalised metric we thus collect the terms to find
\ba
\label{redmet}
{\cal M}_{M N} &=& e^{-\g} \left(  \begin{array}{cc}
    h - b h^{-1}b  &    \frac{-1}{\sqrt{2}} b h^{-1} \\ 
    \frac{1}{\sqrt{2}} h^{-1}b &  \frac{1}{2} h^{-1} \\ 
  \end{array} \right)  \\
&& \qquad + e^{+2 \g} \left(  \begin{array}{cc}
   A_{n}A_{m}\Lambda + \frac{1}{2} K^{2}_{m n }  +\frac{1}{2}A_{m}K_{n} +\frac{1}{2}A_{n}K_{m}  &    - \frac{1}{\sqrt{2}}  ( A_m A^p b_p{}^n +  K_{m p}{}^nA^p  ) \\ 
   - \frac{1}{\sqrt{2}}  ( A_n A^p b_p{}^m +  K_{n p}{}^m A^p  ) & \frac{1}{2}  \left(A^2  h^{nl} - A^l A^n \right) \nonumber
  \end{array} \right)
\ea
and
\be
{\cal A}_{M}=e^{ 2\gamma}   \left(  \begin{array}{c } 
 A_n \Lambda  +  \frac{1}{2} K_{n  }  \\
    -  \frac{1}{\sqrt{2}}  A^m b_m{}^n 
\end{array} \right)  \ , \quad {\cal B}_{M}^{rs}=e^{ 2\gamma}   \left(  \begin{array}{c } 
  \frac{1}{\sqrt{2}}   \left( K_{m}{}^{r s} + A_m b^{r s} \right)   \\
\frac{1}{2} \left(A^s  h^{r m} - A^r h^{s m} \right)
   \end{array} \right)
\ee
We also define 
\be
\tilde{ {\cal B}}_{M} =  {\cal B}_{M}^{rs}b_{rs} =  e^{ 2\gamma}   \left(  \begin{array}{c } 
  \frac{1}{\sqrt{2}}   \left( K_{m}  + A_m b^{2} \right)   \\
-   \left(A\cdot b  \right)^{m}
   \end{array} \right) \ . 
\ee
The goal now is to try and express the second piece of \eqref{redmet} in terms of these `gauge' fields.   Using  that  
\be
\left({\cal A}_{M} - \frac{1}{\sqrt{2}}\tilde{{\cal B}}_{M}  \right) =  e^{2 \g} \left(  \begin{array}{ c} A_m \\ 0
 \end{array} \right) \ ,
 \ee
we find 
\be
{\cal M}_{MN} = e^{{-\g}} {\cal H}_{ M N } +e^{ -2 \g }   \left({\cal A}_{M} - \frac{1}{\sqrt{2}}\tilde{{\cal B}}_{M}  \right)\left({\cal A}_{N} - \frac{1}{\sqrt{2}}\tilde{{\cal B}}_{N}  \right)
 +  e^{{-2 \g}} { \cal  B}_{M}^{r s}{ \cal  B}_{M r s}  \, .
\ee
where $\H_{MN}$ is the usual $O(3,3)$ coset metric\footnote{The peculiar factors floating around in here could have been removed by a rescaling of the dual $y_{ab}$ coordinates, however, we don't do this so as to keep with the conventions of  \cite{Berman:2010is} . }: 
\be
\H_{{MN}} = \left(\begin{array}{cc}
   h - b h^{-1}b  &    \frac{-1}{\sqrt{2}} b h^{-1} \\ 
    \frac{1}{\sqrt{2}} h^{-1}b &  \frac{1}{2} h^{-1}
 \end{array} \right)
\ee  

This structure is somewhat replicates what one finds in a standard KK reduction.   This can be made more explicit if we introduce Greek indices to denote the internal coordinates i.e. $X^{\hat{M}} = (X^M, X_\a) = (X^M , z, y_{ab})$ and define the metric on the internal space 
\be
{\cal G}_{\a \b } =  e^{-2 \gamma}  \left(  \begin{array}{cc}
M_{zz} & M_{z}{}^{r s} \\
  M^{pq}{}_{z} & M^{p q , r s}  
 \end{array} \right)  =  \left(  \begin{array}{cc}
 \Lambda   & \frac{1}{\sqrt{2}}    b^{r s }  \\
\frac{1}{\sqrt{2}}   b^{p q}  &  \frac{1}{2}  \left(h^{p r }h^{s  q} - h^{r q} h^{p s} \right)  
 \end{array} \right) 
\ee
with inverse\footnote{See the appendix for an explanation of how to treat antisymmetric indices to establish the correct factors in this inverse.}  given by 
\be
{\cal G}^{\a \b } =   \left(  \begin{array}{cc}
 1 & - \frac{1}{\sqrt{2}} b_{p q} \\ 
 - \frac{1}{\sqrt{2}} b_{r s}   &  \frac{1}{2}  (h_{p r }h_{s  q} - h_{r q} h_{p s} +b_{p q} b_{r s})
 \end{array} \right) \ .
\ee
 
 Defining the `gauge field' to be
 \be
 \label{Cfield}
{\cal C}_{M \a} = e^{-2 \gamma}   ( {\cal A}_{M}, {\cal B}_{M}{}^{rs} ) \ ,
\ee
  finally allows us to make sense of the KK ansatz applied to the generalised metric.  We find \eqref{Mgenrelabel}  may be written in the form 
 \be
 \label{MKK}
 M_{\hat{M}\hat{N}} =    \left( \begin{array}{c  c}  
e^{-\g} {\cal H}_{M N} + e^{2 \g} {\cal C}_{M \a} {\cal G}^{\a \b }{\cal C}_{N \b}  & \quad e^{ 2 \g}   {\cal C}_{M \a} \\
  e^{ 2 \g}  {\cal C}_{N \b}  & \quad  e^{2 \g}{\cal G}_{\a \b }
 \end{array} \right)  \ . 
 \ee
To find the inverse of $M$ is  easy since the generalised metric is just of a standard KK type: 
\be
M^{\hat{M}\hat{N}}=  (M_{\hat{M}\hat{N}} )^{-1} = \left( \begin{array}{c  c}  
e^{ \g} {\cal H}^{M N}    & \quad - e^{  \g}   {\cal C}^{M \a} \\
 - e^{ \g}  {\cal C}^{N \b}  & \quad  e^{-2 \g}{\cal G}^{\a \b } + e^{  \g} {\cal C}^{P \a}{\cal C}_{P}{}^\b 
 \end{array} \right)
 \ee
in which we have raised indices on the gauge field with either ${\cal H}$ or ${\cal G}$.

   To summarise, the main lesson one can draw from all this is:
\emph{
The standard dimensional reduction gets promoted to a  doubled KK Reduction}.\footnote{Apologies for the name - we hope potential confusion is avoided with a double dimensional reduction of a world sheet - something we don't do in these notes.}  In other words we simply invoke a KK type ansatz on the generalised metric.   In some sense life is simplified through the generalised metric since the  RR fields are entirely contained in the gauge fields of this ansatz.   An important difference however is that the metric on the internal space is \emph{not} independent of the field content of the remaining dimensions.  We shall shortly return to the interpretation of this internal metric but for the meantime we continue directly.  

\subsection{The reduction} 
To perform the dimensional reduction we simply need to plug \eqref{MKK} into \eqref{TBP} and \eqref{VBP} and enforce the vanishing of derivatives in internal directions.   An instructive example of the sorts of manipulations involved is  
\ba
\label{MidM}
 M^{\hat{K}\hat{L}} \partial_{M} M_{\hat{K}\hat{L}} &=&  {\cal H}^{{K L}}\partial_{M} {\cal H}_{K L} + {\cal G}^{{\a \b}}\partial_{M} {\cal G}_{{\a \b}} + ( 2 \delta^{\a}_{\a} - \delta^{N}_{N}) \partial_{M}\gamma \nonumber \\
&=& {\cal H}^{{K L}}\partial_{M} {\cal H}_{K L} + {\cal G}^{{\a \b}}\partial_{M} {\cal G}_{{\a \b}} + 2 \partial_{M}\gamma \, . 
\ea
Notice how in this calculation gauge fields cancel -- this is something that will happen repeatedly in the dimensional reduction.  We now need to make use of a few identities, which may be shown by a brute force calculation, 
 \be
 \label{keyids}
 \H^{{KL}}\partial_{M}\H_{KL} = 0  \, \quad   {\cal G}^{{\a \b}}\partial_{M} {\cal G}_{{\a \b}} =  \Gamma_{M} =  - 2 h^{{rs}} \partial_{M}h_{r s} \ ,
 \ee 
 which allow us to conclude that 
 \be
  M^{\hat{K}\hat{L}} \partial_{M} M_{\hat{K}\hat{L}}= 2 \pd_M \g  +\Gamma_M \ . 
 \ee
 Let us get slightly ahead of ourselves and mention that the term $(h^{-1} \pd_M h)$ is crucial to the definition of the T-duality invariant dilaton and can only arise in this dimensional reduction through the derivatives of the internal metric $\G$.

For the first term in the potential we find:
 \ba
 \label{V1}
V_{1} &=&   \frac{1}{12}  e^{{\g}}\left(  \partial_{M} \H_{KL} \partial^{M}\H^{KL}  -  22  \partial_{M}\g  \partial^{M}\g    +    \partial_{M} \G_{\a \b} \partial^{M} \G^{\a \b} - 4  \Gamma_{M}
   \partial^{M} \g   \right) \nonumber \\
 && \quad- \frac{1}{6} e^{{4\g }}\left( \G^{\a\b} \H^{KL} \partial_{M} \C_{K\a} \partial^{M}C_{L\b}   - 2 \C^{K \a} \G^{\b \gamma} \partial_{M} \C_{K \b} \partial^{M}G_{\a \gamma}  
 + \C_{K}{}^{\a}\C^{K \b} \G^{\s \rho}  \partial_{M} \G_{\a \s} \partial^{M} \G_{\b \rho} \right) \, . \nonumber \\
\ea
The structure of the KK ansatz ensures that a great many of the possible terms in the gauge fields cancel out in these contractions and we can see that  the dilaton scaling is homogenous in the gauge field sector and metric sector respectively.  
In a similar fashion, making use of \eqref{MidM} and the identities \eqref{keyids} one eventually finds
\ba
V_{3}& =& \frac{1}{3}e^{\g} \left( \partial_{M }\gamma +   \frac{1}{2} \G_{M} \right)^{2}  \ ,  \\ 
V_{2}&=& \frac{1}{2}e^{\g} \left( -\H^{KL} \pd_{K} \H^{QN} \pd_{N} \H_{LQ}  +2 \pd_{K}\H^{KN} \pd_{N} \g + \pd_{M} \g  \pd^{M}\g \right)   \\
&& + \frac{1}{2} e^{4\g} \left( \G^{\a \b} \pd^{K} \C_{M \a} \pd^{M}\C_{K \b} - 2 C_{K}{}^{\b}\G^{\a \rho } \pd^{K} \C_{M \a} \pd^{M} \G_{\b \rho} 
+  \C^{M\g} C^{N \mu} \G^{\rho \sigma} \pd_{M}\G_{\mu \rho} \pd_{N}\G_{\g \sigma} \right) \ , \nonumber\\
V_{4} &=& -\frac{1}{2} e^{{\g}} \left(  \pd_{K} \H^{KL} \pd_{L}\g   + \pd_{M} \g \pd^{M} \g  +  \frac{1}{2} V_{K} \pd_{M} \H^{K M} + \frac{1}{2} V_{K} \pd^{K}\g\right)\ .
\ea

The kinetic terms are evaluated in the same way and are simply given by the expression $V_1 + V_3$ with the contracted derivatives replaced by time derivatives.

 \subsection{The NS sector}
Let us first concentrate on the would-be NS sector by temporarily setting the KK gauge fields $\C_{M \a}$ to zero.    To proceed we note some crucial identities that allows the quadratic term of the internal metric to be written in terms of the external metric:
\be
\pd^M\H^{KL}\pd_M \H_{KL}  = 2 tr \left( h^{-1}  \pd_M b  h^{-1} \pd^M b -h^{-1}  \pd_M h  h^{-1} \pd^M h   \right) \ , 
\ee
and
\ba
\label{GtoHid}
\pd^M\G^{\a \b} \pd_M \G_{\a \b} &=& tr \left( h^{-1}  \pd_M b  h^{-1} \pd^M b -h^{-1}  \pd_M h  h^{-1} \pd^M h   \right)  -  tr( h^{-1} \pd_M h )^2\\
& =&\frac{1}{2} \pd^M\H^{KL}\pd_M \H_{KL} -\frac{1}{4} \Gamma_M \Gamma^M \ .
\ea
For the first terms in the potential, \eqref{V1}, we may use these identities to find 
\ba
\label{V1red}
V_{1} &=&   \frac{1}{12}  e^{{\g}}\left( \frac{3}{2}  \partial_{M} \H_{KL} \partial^{M}\H^{KL}  -  22  \partial_{M}\g  \partial^{M}\g    - 4  \Gamma_{M}
   \partial^{M} \g   -\frac{1}{4} \Gamma_M \Gamma^M    \right) \ . 
   \ea
Notice how the coefficient of the $\H$ quadratic term receives contributions from both the internal and external metric terms -- this is prototypical of what will happen in general.

Adding all the contributions in the NS sector we find
\ba
\label{VBPred}
V_{BP}& =&  \sqrt{h}   e^{{\frac{\g}{2} }} \left( \frac{1}{8}\pd_M \H^{KL} \pd^M \H_{KL} - \frac{1}{2}\H^{KL} \pd_{K} \H^{QN} \pd_{N} \H_{LQ}     \right)  \nonumber\\
&& + \sqrt{h}  e^{{\frac{\g}{2} }}    \left( -\frac{3}{2} \pd_M \g  -\frac{1}{4} \Gamma_M \pd^M \g  + \frac{1}{16} \Gamma_M \Gamma^M +\frac{1}{2} \pd_M \H^{MN} \left( \pd_N \g - \frac{1}{2}\Gamma_N \right) \right)  \  
\ea 
in which the overall  prefactor received a contribution $e^{{-\frac{\g}{2}}}$ from the determinant in addition to the factor found in \eqref{V1red}.    One might hope that some linear combination of $\pd_M \g$ and $V_M$ could be defined so as to completely simplify the terms in the second line however that proves not to be possible. 

The kinetic terms are found to be
\be
\label{TBPred}
T_{BP}  =  - \sqrt{h}       e^{{- \frac{\g}{2} }}  \left( \frac{1}{8} tr( \dot{\H}^{-1} \dot{\H} ) -\frac{3}{2} \dot{\g}^2 +\frac{1}{16}  \Gamma_t^2  \right)  \ ,
\ee
in which $ \Gamma_t = -2 tr(h^{-1} \dot{h}) $.

\subsection{Comparison to DFT}
 The Lagrangian for DFT is given by
\be
\label{DFT}
S_{DFT}= \int dx d\tilde{x} e^{-2 d} \left(\frac{1}{8} \H^{ M N } \pd_{M} \H^{KL} \pd_{N} \H_{KL} - \frac{1}{2} \H^{M N}\pd_{M}\H^{KL} \pd_{L} \H_{KN} -2 \pd_{M} d \pd_{N} \H^{MN}
+ 4\H^{MN} \pd_{M} d \pd_{N} d  \right)\ . 
\ee
 The duality invariant DFT dilaton is related to the usual dilaton according to 
\be
e^{-2 d} = \sqrt{|g|} e^{-2 \phi } \ .  
\ee
In DFT \eqref{DFT} all coordinates are doubled (including time) and whereas in the preceding M-theory treatment not only was time not doubled but covariance was broken (by choosing synchronous gauge) and time treated separately.\footnote{To avoid introducing an extra alphabet we have abused notation  in   \eqref{DFT}  such that capital Roman indices run over both doubled spatial  ($x^i$ and  $\tilde{x}_i$)  and doubled temporal coordinates ($t$ and $\tilde{t}$).  Everywhere else the  capital Roman indices run over doubled spatial coordinates only.}  The first thing we do is to mirror this by separating the potential and kinetic terms from \eqref{DFT} choosing the off-diagonal pieces $\H_{t I} = \H_{\tilde{t} I} = \H_{t \tilde{t}}  = 0$ and by setting $\frac{\partial}{\partial  \tilde{t}} \equiv 0$.  Using $\H_{tt} = \H_{\tilde{t} \tilde{t}}^{-1}$ we find that 
\ba
\label{TDFT}
T_{DFT} &=&  e^{-2 d} \left( \frac{1}{8}\H^{tt} \dot{\H}^{MN} \dot{\H}_{MN} - \frac{1}{4} \dot{\H}^{tt} \dot{\H}_{tt} -2 \dot{d} \dot{\H}^{tt} +4 \dot{d}^2 \H^{tt}   \right)  \ , \\ 
\label{VDFT}
V_{DFT} &=&  e^{-2 d} \left(\frac{1}{8} \H^{ M N } \pd_{M} \H^{KL} \pd_{N} \H_{KL} - \frac{1}{2} \H^{M N}\pd_{M}\H^{KL} \pd_{L} \H_{KN} \right. \nonumber \\ 
&& \qquad \left. -2 \pd_{M} d \pd_{N} \H^{MN} + 4\H^{MN} \pd_{M} d \pd_{N} d +\frac{1}{4}\H^{MN}\pd_M \H^{tt} \pd_N \H_{tt}   \right) \ .
\ea

Now we must understand what the correct choice is for $\H_{tt} = g_{tt} $ in order to make contact with the M-theory reduction.  This is slightly subtle,  when reducing from eleven to ten dimensions as in section 2 of this paper one performs a Weyl rescaling in all directions, including time.  However, in the Berman-Perry approach,  the time component of the M-theory metric $G_{tt}$ had been gauge fixed to unity.  Thus to make contact with the preceding calculations we must constrain the  time component of the string theory metric $g_{tt}$ to obey 
\be
-1 = G_{tt} = e^{-\g} g_{tt} \Rightarrow g_{tt} = \H_{tt} = e^\g  \ , 
\ee
and the double dilaton to obey
\be
e^{-2 d} = \sqrt{h} e^{-2 \phi + \frac{\gamma}{2} } \ .  
\ee

Let us compare the  dilaton prefactors arising in the reduced potential  with that in the DFT.   We shall do this in more generality than the case at hand by including an arbitrary number of dimensions. Consider the M-theory defined with $n$ spatial dimensions so that the corresponding dimensional reduction has  $n - 1$ spatial dimensions.  Under the standard KK ansatz, \eqref{kk1}, the determinant of the M-theory spatial metric is given by $\det{G} = e^{(3 - n)\g} \det{h} $.   Thus the reduced M-theory potential is schematically given by   
\be
V_{BP} \approx \sqrt{G} M^{\hat{M}\hat{N}} \partial_{\hat{M}} M^{\hat{K}\hat{L}}\partial_{\hat{N}} M_{\hat{K}\hat{L}} + \dots  \approx   e^{(3-  n)\frac{\g}{2}} \sqrt{h}  e^{\g}  \H^{ M N } \pd_{M} \H^{KL} \pd_{N} \H_{KL} +\dots 
\ee
where the extra factor of $ e^{\g}$ comes from the contraction of indices with $M^{-1}$  (for  $n = 4$ we thus have an overall prefactor  $e^{\frac{\g}{2} }$ as in eqn.  \eqref{VBPred} ).  To match this to the prefactor in the DFT potential, \eqref{VDFT}, we must have 
\be
 e^{(3- n)\frac{\g}{2}} e^{\g} \sqrt{h} = \sqrt{h} e^{-2 \phi + \frac{\gamma}{2} }
\ee
thus
\be
\left( 2 - \frac{n}{2} \right) \g = - 2\phi \ , 
\ee
which correctly reproduces the relation between KK scalar and dilaton field found in \eqref{gammadilaton}.  In  comparing the prefactors of the kinetic terms  one arrives at the same conclusion. 

Therefore, as anticipated in section 2, although the dimensionally reduced theory is perfectly valid in its own right, in order for it to be compared with DFT we should set the standard dilaton $\phi$ and $\gamma$ both to zero.  This does not imply that the doubled dilaton is zero; instead we have $d = - \frac{1}{4} \ln \det h$  and for its derivatives
\be
\pd_{M} d = - \frac{1}{4} tr(h^{-1} \pd_{M} ) = 8 \Gamma_{M}  \ .
\ee
With this it is immediately clear that the dimensional reduction agrees with the DFT. 

One might perhaps be able to alter the formulation of the DFT to match excatly with the dimensionally reduced M-theory however there seems relatively little to be gained in doing so since we know the DFT is equivalent to string theory.  Instead, a more useful enterprise is surely to develop a completed duality invariant M-theory in $1+10$ dimensions.

\subsection{The RR sector}
Let us now turn to the gauge fields.  Combing the contributions to the potential yields:
\ba 
\frac{e^{-4\g}}{\sqrt{G} } V_{C}& =&    - \frac{1}{6}  \, {\G}^{\alpha \beta} {\H}^{K L} {\partial}_{M}{{C}_{K \alpha}}\,  {\partial}^{M}{{\C}_{L \beta}}  -\frac{1}{6}  {\C}_{K \alpha} {\C}_{L \beta} {\G}^{\alpha \gamma} {\G}^{\beta \mu} {\G}^{\rho \sigma} {\H}^{K L} {\partial}_{M}{{\G}_{\gamma \rho}}\,  {\partial}^{M}{{\G}_{\mu \sigma}}
\nonumber  \\&& + \frac{1}{3}   {\C}_{K \alpha} {\G}^{\alpha \beta} {\G}^{\gamma \rho} {\H}^{K L} {\partial}_{M}{{\C}_{L \gamma}}\,  {\partial}^{M}{{\G}_{\beta \rho}}    + \frac{1}{2}\, {\G}^{\alpha \beta} {\partial}^{K}{{\C}_{L \alpha}}\,  {\partial}^{L}{{\C}_{K \beta}}   \, \nonumber   \\
 &&  - {\C}_{K \alpha} {\G}^{\alpha \beta} {\G}^{\gamma \rho} {\partial}^{K}{{\C}_{L \gamma}}\,  {\partial}^{L}{{\G}_{\beta \rho}}\,   + \frac{1}{2}\, {\C}_{K \alpha} {\C}_{L \beta} {\G}^{\alpha \gamma} {\G}^{\beta \mu} {\G}^{\rho \sigma} {\partial}^{K}{{\G}_{\mu \rho}}\,  {\partial}^{L}{{\G}_{\gamma \sigma}}\,   \nonumber \\
\ea

This can be simplified by defining a new derivative
\be
D_M \C_{K \a} = \pd_M \C_{K \a}  -\C_{K \b}   ( \G^{\b \s} \pd_M \G_{\s \a} ) \ , 
\ee
to yield 
\be
\frac{e^{-4\g}}{\sqrt{G} } V_{C} = - \frac{1}{6} \G^{\a \b} D_M \C_{K \a}  D_N \C_{L \b}  \left(\H^{MN} \H^{KL} - 3 \H^{MK} \H^{NL}  \right)
\ee
The relative factor between the two pieces means that this combination is not a standard field-strength-squared contraction.    Whilst this result is relatively simple, the field $\C_{M \a}$ defined according to \eqref{Cfield} is rather difficult to interpret; it contains various contractions of the RR potentials together with the metric and NS two-form.   This can be remedied by noticing that upon raising an internal index on the gauge field we have
\be
\label{cform}
\C_{M}^\a = \G^{\a\b} \C_{M \b} = \left(\begin{array}{cc}  A_m  & \frac{1}{\sqrt{2}}K_{m r s} \\
0 & \mathbb{1}^{mn}_{rs} A_n    \end{array}\right) \ , 
\ee
in which $\mathbb{1}^{nm}_{rs} = \frac{1}{2} \left(\d^n_r \d^m_s - \d^m_r \d^n_s \right)  $ is the appropriate identity operator for antisymmetric indices (see appendix).    In this form we see that the NS dependence has dropped out entirely.   Furthermore, the derivative introduce above is no more than 
\be
D_M \C_{K \a} = \G_{\a \b}\pd_M \C_K^\b
\ee
and the RR sector of the dimensionally reduced theory is very simply given by 
\be
\label{RRterms}
\frac{e^{-4\g}}{\sqrt{G} } V_{C} = - \frac{1}{6} \G_{\a \b} \pd_M \C_{K}^\a  \pd_N \C_{L}^\b  \left(\H^{MN} \H^{KL} - 3 \H^{MK} \H^{NL}  \right) \ . 
\ee
Upon enforcing that $\frac{\partial}  {\partial \tilde{x}} = 0$ and expanding out the above expression one indeed recovers exactly the correct contributions to the two-form and four-form RR field strengths found in IIA supergravity.    The calculation is a little laborious but straightforward  and again relies on a delicate interplay between the internal metric $\G_{\a\b}$ and the   metric $\H_{IJ}$. \footnote{To compare with certain texts it is important to keep in mind our definition of $A^{(3)} = K^{(3)}  -A^{(1)} \wedge b$ hence the modified field strength $\tilde{F}^{(4)} = dA^{(3)} - A^{(1)}\wedge H = dK^{(3)} - dA^{(1)} \wedge b$. It is for this reason that contractions in \eqref{RRterms} carry no derivatives of the two-form $b$.  } 

\subsubsection{Interpreting the internal metric and the KK gauge fields} 
At this stage we consider further what is the meaning of the internal index structure on these gauge fields.  In a standard KK reduction the internal metric does not have any relation to the external metric. The isometries of the internal metric translate to the gauge symmetries of the gauge field.  However the situation is quite different here; the internal metric depends on the same functions (i.e. the components of the real metric $h$ and NS two-form $b$) as the external metric $\H$ but in a different representation of the $O(d,d)$ symmetry group.    How then should we really interpret the internal metric?  A key observation is the isomorphism $Spin(3,3) \cong Sl(4, \mathbb{R} )$.   It is thus natural to consider the internal metric (which we recall acts in a four-dimensional space) as a spinorial counter part to the external metric $\H$.    Since $\H$ describes an $O(3,3)/O(3)\times O(3)$ coset, one might expect that the internal metric defines a $Sl(4, \mathbb{R} ) /SO(4)$ coset space. This is not quite accurate;  the internal metric is not unimodular but instead has a determinant equal to $\det(h)^{-2}$.\footnote{This result seems to mirror a subtlety concerning spinors in   generalised geometry: the spin bundle is actually $S = \Lambda^{\bullet}T^{\ast} \otimes det(T)^{\frac{1}{2}}$ and the action of $Gl(d)$ on spinors is modified.}  To see this more explicitly the first thing to do is to remove the rather awkward antisymmetric indices on the internal coordinates.  This is achieved by defining $ z\equiv  z^{0}$ and  $ y_{{ij}} =     \frac{1}{\sqrt{2}} \eta_{{ij k}} z^{k}$ where $\eta_{ijk}$ is a permutation symbol (taking values $0,\pm 1$) .    Then the internal line-element
\be
ds^{2}_{int} =\G_{\a \b} dy^{\a} dy^{\b} = ( 1+ \frac{1}{2} b_{pq}b^{pq}  ) dz^0 dz^{0} + \frac{2}{\sqrt{2}} b^{pq} dz dy_{pq} + \frac{1}{2}(h^{p r} h^{q s} - h^{p s} h^{q r} ) dy_{pq} dy_{rs}  \ , 
\ee
becomes 
\ba
ds^{2}_{int}  
&=&  ( 1+   \b_{k}\b^{k}  ) dz^0 dz^{0} +   2  \b_k \det{h}^{{-1/2}}  dz dz^{k} +      h_{ij} \det{h}^{{-1}}  dy^{i}  dy^{j} \ , 
\ea
in which  we have defined     $2 \b_i =  b^{ij} \e_{i j k }$  ($\e$ being the Levi-Civita antisymmetric tensor).

We remark that the $SL(4)/SO(4)$ coset has appeared previously in the literature;   the reduction of five-dimensional dilaton-axion gravity on a $T^2$ can be formulated with such a coset \cite{Kechkin:1997yj} as can the  KK reduction of six-dimensional pure gravity on a $T^3$ \cite{Cvetic:1995sz}. 

With this in mind it now seems a possible interpretation of the KK guage fields $\C_{A   }^\a$  is that they are bosonic spin 3/2 fields of the double field theory.  As is the case with the encoding of the NS sector fields in $\H$, these $\C_{A}^\a$ are not free but are constrained by the form of \eqref{cform}.   Should this result hold true in general,   one consequence of this is that the forms of both the field $\C$ and the spinorial metric $\G$ will vary according to the dimensionality; this is in contrast to $\H$ which takes the form regardless of dimension.    We do not rule out the possibility that there may be an alternative, and more universal, formulation that makes use of   $\H$ alone  but this certainly seems rather unnatural  given perspective gained from the preceding M-theory considerations .

\section{From $O(5,5)$ covariant M-theory to $O(4,4)$ DFT} 

We now turn to the next dimension up in which we start with a supergravity theory with five spatial dimensions.  In this case, through the introduction of eleven extra coordinates it is possible to recast the theory in a way that displays manifest $SO(5,5)$ invariance \cite{Berman:2011pe}.   Note that the coordinates describe a sixteen of $SO(5,5)$ and \emph{not} the vector of this group (as would be the case in the DFT).  Due to this the form of the generalised metric is rather more involved.  The sixteen coordinates are now $X^{\hat{M} } =   \{X^{\hat{a} }, Y_{\hat{a} \hat{b}} , U  \}$ and the generalised metric is given by
\be
M_{\hat{I}\hat{J} } = \left( \begin{array}{c c c} 
G_{\hat{a}\hat{ b}} + \frac{1}{2} C_{\hat{a} \hat{e}\hat{f}} C_{\hat{b}}{}^{\hat{e} \hat{f}} +\frac{1}{16}X_{\hat{a}} X_{\hat{b}}  & \frac{1}{\sqrt{2}} C_{\hat{a}}{}^{\hat{k}\hat{l} } +  \frac{1}{4\sqrt{2}}X_{\hat{a}} V^{ \hat{k}\hat{l} } &  \frac{1}{4}G^{-1/2} X_{\hat{a}}   \\ 
 \frac{1}{\sqrt{2}} C_{\hat{b}}{}^{\hat{m}\hat{n}} +  \frac{1}{4\sqrt{2}}X_{\hat{b}} V^{ \hat{m}\hat{n} }&  \frac{1}{2}(G^{\hat{m}  \hat{k}} G^{\hat{n}\hat{l}}  -G^{\hat{m}  \hat{l}} G^{\hat{k}\hat{n}})    + \frac{1}{2} V^{ \hat{k}  \hat{l}} V^{ \hat{m}  \hat{n}}  &  \frac{1}{\sqrt{2} } G^{-\frac{1}{2}}   V^{ \hat{m}  \hat{n}}  \\
  \frac{1}{4}G^{-1/2} X_{\hat{b}} &   \frac{1}{\sqrt{2} } G^{-\frac{1}{2}}   V^{ \hat{k}  \hat{l}}   &G^{-1}
 \end{array} 
     \right)
\ee
in which
\ba
V^{\hat{a}\hat{b} } = \frac{1}{6} \e^{\hat{a}\hat{b} \hat{c}\hat{d} \hat{e} } C_{\hat{c} \hat{d}\hat{e}} \ , \quad
X_{\hat{a}}= V^{\hat{d} \hat{e}} C_{\hat{d} \hat{e} \hat{a}}  \ , \quad   G = \det G \ ,
\ea
where the epsilons are tensors (not densities) defined according to 
\be
\e^{\hat{a}\hat{b} \hat{c}\hat{d} \hat{e} }  = \frac{1}{\sqrt{G}} \eta^{ \hat{a}\hat{b} \hat{c}\hat{d} \hat{e}  } \ , 
\ee
where $\eta$ is an alternating permutation symbol taking values $\{-1,0,1\}$.   The inverse metric is given by 
\be
M^{\hat{I}\hat{J} } =\left(\begin{array}{ccc} 
G^{\hat{a}\hat{b} } & -\frac{1}{\sqrt{2}} C^{\hat{a} }{}_{\hat{m}\hat{n} } & \frac{\sqrt{G}}{4} X^{\hat{a}} \\ 
 -\frac{1}{\sqrt{2}} C^{\hat{b} }{}_{\hat{p}\hat{q} }  & \quad G_{\hat{p}\hat{q},\hat{ m}\hat{n} } + \frac{1}{2}  C_{\hat{p} \hat{q}}{}^{\hat{a}}C_{\hat{a} \hat{m} \hat{n}} & -\frac{\sqrt{G}}{\sqrt{2} }  V_{\hat{p} \hat{q}} - \frac{\sqrt{G}}{4 \sqrt{2} }   C_{\hat{p} \hat{q} \hat{a}} X^{\hat{a}} \\ 
\frac{\sqrt{G}}{4} X^{\hat{b}} & -\frac{\sqrt{G}}{\sqrt{2} }  V_{\hat{m} \hat{n}} - \frac{\sqrt{G}}{4 \sqrt{2} }   C_{\hat{m} \hat{n} \hat{a}} X^{\hat{a}} & \quad   1 + \frac{1}{2}V_{\hat{a} \hat{b}} V^{\hat{a} \hat{b}}   + \frac{1}{16} X^{\hat{a}}X_{\hat{a} } 
\end{array} \right) \ . 
\ee

The structure of the Hamiltonian in this case is the same as in the $SL(5)$ case in  \eqref{TBP},\eqref{VBP} but with different coefficients for each term; the kinetic terms are 
\be
T_{BP} = -\sqrt{G} \left(  \frac{1}{16} tr (\dot M^{-1} \dot M ) + \frac{3}{128} (tr (  M^{-1} \dot M ))^2 \right)
\ee
and the potential is given as
\be
\frac{1}{\sqrt{\det{G}} } V_{BP} = V_1  + V_2 + V_3 + V_4 
\ee
with 
\ba
V_1 &=& \frac{1}{16} M^{\hat{M}\hat{N}} (\partial_{\hat{M}} M^{\hat{K}\hat{L}}) (\partial_{\hat{N}} M_{\hat{K}\hat{L}}) \ , \\
V_2 &=&- \frac{1}{ 2} M^{\hat{M}\hat{N}} (\partial_{\hat{N}} M^{\hat{K}\hat{L}}) (\partial_{\hat{L}} M_{\hat{K}\hat{M}}) \ , \\
V_3 &=&  \frac{3}{128} M^{\hat{M}\hat{N}} ( M^{\hat{K}\hat{L}} \partial_{\hat{M}} M_{\hat{K}\hat{L}})( M^{\hat{R}\hat{S}} \partial_{\hat{N}} M_{\hat{R}\hat{S}})\ , \\ 
V_4 &=&   \frac{1}{8} M^{\hat{M}\hat{N}}  M^{\hat{P}\hat{Q}}  ( M^{\hat{K}\hat{L}} \partial_{\hat{P}} M_{\hat{K}\hat{L}})(   \partial_{\hat{M}} M_{\hat{N}\hat{Q}})\ . 
\ea 
 At first glance the coefficients in these expressions seem rather strange, and it is  certainly not obvious how they should relate to the coefficients found in the DFT.  However, we shall see that they are, of course, exactly what is needed to ensure delicate cancelations occur.  

\subsection{KK ansatz in the NS sector }
Now we make the split of the sixteen into the eight internal coordinates as $X^{\a} =\{ U, z, Y_{a b}\} $ and the external coordinates as $X^I = \{X^a , Y_{z a} \}$.   First let us just consider the would-be NS sector i.e. we have 
\be
G_{a b } = e^{-\g} h_{a b} \ , \quad G_{z z} = e^{ 2 \g} \ , \quad C_{a b z} = b_{a b} \ , 
\ee
and all other components zero.    Also we define the lower-dimensional epsilon tensor as 
\be
\e^{ a b c d }  = \frac{1}{\sqrt{h}} \eta^{ a b c d} \ ,
\ee
noting that the relationship between the five- and four-dimensional tensors is
\be 
\e^{ a b c d  z} =  \frac{1}{\sqrt{G}} \eta^{  a b c d z  } = e^\g \frac{1}{\sqrt{h}} \eta^{  a b c d   } = e^\g \e^{ a b c d }  \ .
\ee
Under this decomposition we have 
\be
V^{z a}  = 0 \ , \quad V^{a b} = \frac{1}{2} e^{  \g} \e^{a b c d} b_{c d} \ , \quad X_z = \frac{1}{2} e^{  \g}   \e^{a b c d} b_{a b} b_{c d} \ , \quad X_a = 0 \ , \quad \det G = e^{-2 \g} \det h. 
\ee

The generalised metric then splits according to 
\be
M_{\hat{I}\hat{J} }  =\left( \begin{array}{cc} 
e^{-\g} \H_{I J}& 0 \\
0 & e^{2 \g} \G_{\a \b}
 \end{array}\right)
\ee
where $\H_{I J}$ is the $O(4,4)$ generalised metric and the internal metric is given by
\be
 \G_{\a \b}
 = \left( \begin{array}{ccc} 
 h^{-1} & \frac{1 }{8\sqrt{h} }\Theta & \frac{ 1 }{2 \sqrt{2 h }} \beta^{a b}  \\
    \frac{1 }{8\sqrt{h} }\Theta & 1 + \frac{1}{2} b^2 + \frac{1}{64}  \Theta^2  & \frac{1}{\sqrt{2}} (b^{a b} + \frac{1}{16} \Theta \beta ^{ a b} ) \\
    \frac{ 1 }{2 \sqrt{2 h }} \beta^{c d}  & \quad  \frac{1}{\sqrt{2}} (b^{c d} + \frac{1}{16} \Theta \beta ^{ c d} ) &  \quad   h^{a b,c d}  + \frac{1}{8}  \b^{ a b} \b^{ c d}
 \end{array}\right) \ , 
\ee
with   inverse
\be
\G^{\a \b}
 = \left( \begin{array}{ccc} 
 h \left( 1 + \frac{1}{64} \Theta^2 + \frac{1 }{8}\beta^{ab} \beta_{ab}  \right) & \frac{  \sqrt{h} }{8} \Theta & -\frac{  \sqrt{h} }{2 \sqrt{2} } \left( (\beta_{ab} + \frac{1}{4}\Theta b_{ab}  \right) \\
  \frac{  \sqrt{h} }{8} \Theta  & 1 & -\frac{1}{\sqrt{2}} b_{ab} \\
 -\frac{  \sqrt{h} }{2 \sqrt{2} } \left( \beta_{c d} + \frac{1}{4}\Theta b_{c d}  \right) &  -\frac{1}{\sqrt{2}} b_{c d} & h_{ab , cd} +\frac{1}{2} b_{ab} b_{cd} 
  \end{array}\right) \ . 
\ee
in which $\Theta = \e^{a b c d }b_{a b} b_{c d}$ and $\beta^{ a b } = \e^{a b c d}b_{c d}$. 

In this case we have the following identities for the derivatives of the internal metric: 
\ba
\label{Gid1}
tr( \G \partial \G^{-1} ) &= & 4  tr (h^{-1} \partial h ) \ , \\
\label{Gid2}
tr(\partial \G \partial \G^{-1} )  
&=& -2( tr (h^{-1} \partial h )  )^2 + tr(\partial \H \partial \H^{-1} )  \ . 
\ea
These allow us to determine that  
\ba
\label{mipm}
tr( \M^{-1} \partial  \M )& = & (2 tr( \G^{-1} \G) - tr( \H^{-1} \H )) \pd \g +  tr( \G^{-1} \pd  \G ) \nonumber \\
&=& 32 \pd \phi +16 \pd d 
\ea 
and
\ba
\label{pmipm}
tr(\pd\M^{-1} \partial  \M ) & = & tr \pd \H^{-1} \pd \H +  tr \pd \G^{-1} \pd \G - ( tr( \H^{-1} \H ) + 4 tr( \G^{-1} \G) ) \pd \g \pd \g  - 4 \pd \g   tr( \G^{-1} \pd  \G )  \nonumber \\
&=&  2 tr \pd \H^{-1} \pd \H - (2\cdot 4\cdot 4) \pd d \pd d   -  (40\cdot 16 )\pd \phi \pd \phi  - (4 \cdot 16\cdot 4)  \pd \phi  \pd d \ ,
\ea
in which we have defined (for reasons that will become apparent upon comparing to the DFT) 
\be
\label{dilrel}
\pd d  =  -\frac{1}{4} tr (h^{-1} \partial h )
\ee
and used the relation between the KK scalar and string theory dilaton in this dimension, $\g = 4\phi$. 

Now we substitute this ansatz into the potential to find 
\ba
V_1+ V_3 &=& e^{4\phi} \H^{MN} \left(  \frac{1}{8}   \pd_M \H^{KL} \pd_N \H_{KL}+ 4\pd_M d \pd_N d + 8 \pd_M \phi  \pd_N d - 8 \pd_M \phi \pd_N \phi \right) \\ 
V_2 &=& e^{4\phi} \left(8 \H^{MN} \pd_M \phi \pd_N \phi  +4 \pd_K \H^{KL}\pd_L \phi - \frac{1}{2} \H^{MN}\pd_N\H^{KL} \pd_L \H_{MK}   \right)\\
V_4 &=& - e^{4\phi} \left(  4\pd_M \phi \pd_N \H^{MN} + 2\pd_M d \pd_N \H^{MN}  +16 \pd_M \phi \pd^M \phi + 8 \pd_M \phi \pd^M d \right)
\ea
in which we have eliminated the derivatives of the internal metric according to the identities \eqref{Gid1} and \eqref{Gid2}.    Combining terms, and including the $\sqrt{G}$ pre-factor we find the final result for the potential 
\be
V_{BP} =  \sqrt{h}   \left(  \frac{1}{8}   \pd^M \H^{KL} \pd_M \H_{KL} + 4\pd^M d \pd_M d - \frac{1}{2} \H^{MN}\pd_N\H^{KL} \pd_L \H_{MK}  -  2\pd_M d \pd_N \H^{MN}  - 16 \pd^M \phi \pd_M\phi \right) \ . 
\ee
Note that the overall factor of the dilaton has cancelled from the $\sqrt{G}$ pre-factor and the $ e^{4\phi}$ that comes from contracting the indices in the potential with an $M^{-1}$ .

The kinetic terms follow in a similar way and can be read from $V_{1} + V_{3}$ by simply replacing the spatial derivatives to temporal ones and removing the factor of $e^{{4\phi}}$: 
\be
T_{BP} =  -e^{-4\phi} \sqrt{h} \left(  \frac{1}{8}  tr( \dot \H^{-1} \dot \H)+ 4 \dot d^2 + 8 \dot \phi  \dot d - 8 \dot \phi^2 \right)
\ee

\subsection{Comparison to DFT}
We recall  the Lagrangian for DFT is given by
\be
\label{DFT}
S_{DFT}= \int dx d\tilde{x} e^{-2 d} \left(\frac{1}{8} \H^{ M N } \pd_{M} \H^{KL} \pd_{N} \H_{KL} - \frac{1}{2} \H^{M N}\pd_{M}\H^{KL} \pd_{L} \H_{KN} -2 \pd_{M} d \pd_{N} \H^{MN}
+ 4\H^{MN} \pd_{M} d \pd_{N} d  \right)
\ee
in which not only the spatial but also the temporal coordinate has been doubled.  The duality invariant dilaton is related to the usual according to 
\be
e^{-2 d} = \sqrt{|g|} e^{-2 \phi }
\ee
To make contact with the approach of Berman--Perry we should separate out the time from this action.  Accordingly we split the indices up as $X^{M} = \{X^M , t , \tilde{t} \} $ and us make the following assumptions:
\be
\label{dftansatz}
\H_{K t} = 0 \ , \quad \H_{K \tilde{t} } = 0 \ , \quad \H_{t t} = -e^{  4 \phi} , \quad  \H_{\tilde{t}\tilde{t}} = -e^{- 4 \phi} \ , \quad \partial_{\tilde{t}} = 0 \ . 
\ee
As explained in section 3, these assumptions serve to mirror the gauge fixing choice made in the derivation of Berman--Perry.  The consequence of the Weyl rescaling is that we must use:
\be
-1 = G_{t t} = e^{-\g} h_{t t}  \Rightarrow h_{t t} = - e^{\g}  = - e^{ 4 \phi} \ .
\ee
This also has an implication for the doubled dilaton:
\be
e^{-2 d} = \sqrt{|g|} e^{-2 \phi } = |g_{{tt}}|^{\frac{1}{2}} \sqrt{h}e^{-2 \phi } = \sqrt{h}
\ee
which is in accordance with the definition \eqref{dilrel}.

The Lagrangian \eqref{DFT} thus splits into a kinetic piece given by 
\be
T_{DFT} = -e^{-2 d } e^{-4 \phi} \left( \frac{1}{8} \dot \H^{-1} \dot \H + 4 \dot \phi^2 + 8 \dot d \dot \phi + 4 \dot d^2 \right)
\ee
and potential terms
\ba
V_{DFT} &=& e^{-2 d} \left(\frac{1}{8} \H^{ M N } \pd_{M} \H^{KL} \pd_{N} \H_{KL} - \frac{1}{2} \H^{M N}\pd_{M}\H^{KL} \pd_{L} \H_{KN} -2 \pd_{M} d \pd_{N} \H^{MN}  \right. \nonumber \\ 
&& \qquad \qquad \left.+ 4\H^{MN} \pd_{M} d \pd_{N} d  - 4\H^{MN}\pd_M \phi \pd_N \phi \right) \ .
\ea

 With the exception of the coefficients of the quadratic terms in $\pd \phi$ these are in agreement with the potential and kinetic terms obtained by dimensional reduction.  This discrepancy is something that we anticipated from the outset due to the arguments of section 2 and is something that is unavoidable when working away from the true critical dimensions.   Nonetheless, the terms involving the doubled diltaton, $d$, are correctly reproduced and this was only possible with a careful treatment of the Weyl rescaling effects on time.

\subsection{RR Sector}
The inclusion of the RR fields is much the same as before (though the calculation is rather more involved); the standard KK ansatz produces a Doubled KK ansatz of the form  \eqref{MKK}.   The gauge fields of this KK ansatz contain the dependance on the RR sector and although $\C_{M  \a}$ is a rather unwieldy expression upon raising the internal index with the metric $\G^{\a \b}$ the structure simplifies dramatically to give
\be
\C_{M}^{\a} = \ \left(\begin{array}{ccc}  \frac{\sqrt{h} }{12 } \e^{p k l n}K_{k l n}b_{ m p}  - \frac{\sqrt{h} }{8 } K_{m k l  }\e^{k l p q}b_{p q}  &\quad A_m  & \quad  \frac{1}{\sqrt{2}}K_{m r s} \\
 \frac{\sqrt{h} }{\sqrt{2}}\e^{m k l n}K_{k l n} & 0 & \mathbb{1}^{mn}_{rs} A_n    \end{array}\right) \ .
\ee
Upon dimensional reduction we find that the only terms with RR fields are again   quadratic in derivatives and have the structure
\be
\label{RRterms2}
\frac{e^{-4\g}}{\sqrt{G} } V_{C} = - \frac{1}{8} \G_{\a \b} \pd_M \C_{K}^\a  \pd_N \C_{L}^\b  \left(\H^{MN} \H^{KL} - 4 \H^{MK} \H^{NL}  \right) \ . 
\ee

 \section{Discussion} 
 To summarise, aside from some dimensional dependent subtleties with the dilaton, upon dimensional reduction,  the duality invariant approach to M-theory descends to the duality invariant doubled field theory for the NS fields.    
 
 In addition, due to the democratic way in which M-theory treats the would-be Ramond and NS sectors, this dimensional reduction has provided further perspective on the way RR fields might be incorporated into the DFT. We suggest  that the RR fields should be encapsulated in the form of the KK gauge potentials and carry an external $O(d,d)$ vector index and internal $Spin(d,d)$ fundamental spinor index.  In this way they might naturally be thought of as spin $3/2$ fields in the  DFT.   One might anticipate that this result holds in general for all $O(d,d)$ groups.   For the next dimension up,  the theory with  6 spatial directions, we already encounter some  discrepancy. In that case we expect to formulate the M-theory in an $E_{{6,6}}$ covariant manner by the introduction of twenty one extra coordinates (corresponding to 15 membrane wrapping charges and six fivebrane wrapping charges).   Then upon dimensionally reducing to the $O(5,5)$ T-duality invariant DFT we would have seventeen internal directions.  Under $O(5,5)$ we have the decompostition $\bf{27} \rightarrow \bf{10} + \bf{16}  +\bf{1}$.   Thus,  in this case one would again expect to find a KK gauge field with a vector and spinor index of $O(5,5)$ but also some extra $U(1)$ vector fields due to the singlet in the decomposition.  The interpretation of these, and how they relate to the considerations of  \cite{West:2010ev} will be of interest.  It will also be important to clarify how the exotic local gauge symmetry of DFT acts on these fields (a result that of course should be obtainable from a knowledge of the correct gauge symmetry of duality invariant M-theory).
 
   Should this result hold true in general,   one consequence  is that the forms of both the field $\C$ and the spinorial metric $\G$ will vary according to the dimensionality; this is in contrast to $\H$ which takes the same form regardless of dimension.    We do not rule out the possibility that there may be an alternative, and more universal, formulation that makes use of   $\H$ alone  but this certainly seems rather unnatural  given perspective gained from these M-theoretic considerations. The development of a general treatment of RR fields in the DFT and indeed the incorporation of supersymmetry (which is likely to be rather exotic given the novel gauge symmetry of DFT), remains an area ripe for further exploration.

 To develop this approach to the incorporation of RR into the double field theory it will be necessary to understand the construction of the internal metrics $\G_{a b}$ for general dimension $O(d,d)$ groups.  Actually, beyond just developing the DFT, this would help inform the construction of duality invariant M-theory for larger duality groups.   By the KK ansatz   the correct form of the generalised metric for the M-theory would be readily apparent.   Assuming that the four terms in the M-theory potential remain the same it would then be a simple matter of fixing four coefficients.

 A rather awkward feature that made that obscures the relation between the DFT and the generalised M-theory was the treatment of time.   It would be desirable to form a more covariant version of the generalised M-theory in which time is also in some sense doubled.   Furthermore,  as we have seen the dimensional dependence of the dilaton terms means that the dimensional reduction is not exactly the DFT.  This would be remedied by a treatment that includes all dimensions in the M-theory.  The ultimately, and lofty goal, would be to build an M-theory that exhibits the maximium duality group!

 Finally we remark that there are two (somewhat related) areas where the duality invariant M-theory remains not completely understood: providing a generalisation of the section condition in a duality covariant way and understanding the gauge symmetries of the theory.   One hope is that these might be reverse engineered as lifting of the equivalent constraint and gauge symmetry of the double field theory.

\section{Acknowledgements} 
It is a great pleasure to thank Sung-Soo Kim,   Amitabh Virmani and in particular Neil Copland and David Berman for fruitful discussions surrounding this work.  I thank Peter West and Axel Kleinschmidt for correspondence on the first draft of this manuscript and for pointing out some important references. This work is supported by the Belgian Federal Science Policy Office through the Interuniversity Attraction Pole IAP VI/11 and by FWO Vlaanderen through project G011410N.

 \section*{Note Added}
 After this work appeared as a preprint the following relevant papers appeared on the arXiv:  \cite{Hohm:2011zr, Hohm:2011dv, Coimbra:2011nw}.
 
  
\begin{appendix}

\section{A note on antisymmetric indicies} 
We need the notion of inversion of a matrix whose components cary antisymmetric indices.  To define the inverse we first establish the that the correct identity operator is given by
\be
\mathbb{1}^{ab}_{cd} = \frac{1}{2} \left(\delta^a_c \delta^b_d -   \delta^b_d \delta^a_c  \right)
\ee
which has the properties that 
\be
\mathbb{1}^{ab}_{cd} \mathbb{1}^{cd}_{ef}= \mathbb{1}^{ab}_{ef} \ , \quad \mathbb{1}^{ab}_{cd} T^{cd} = T^{ab}
\ee
when acting on an antisymmetric $T^{ab}$.  Also the trace is given by
\be
\mathbb{1}^{ab}_{ab} = \frac{1}{2} \left( n^2 - n \right)
\ee
correctly counting the dimension of the space of antisymmetric indices.  With this definition the metric 
\be
h^{ab , cd} = \frac{1}{2} \left( h^{a c} h^{b d} - h^{a d} h^{b c} \right)
\ee
has inverse
\be
h_{ab , cd} = \frac{1}{2} \left( h_{a c} h_{b d} - h_{a d} h_{b c} \right)
\ee
properly normalised so that 
\be
h^{ab , cd}h_{cd , ef} = \mathbb{1}^{ab}_{ef} \, . 
\ee

\section{Curvature formulae}
To establish a sign convention we define the Ricci curvature as 
\be
R_{ab}   = \pd_c \Gamma^c_{ab} - \pd_b \Gamma^c_{a c} + \Gamma^c_{ab} \Gamma^d_{c d}  - \Gamma^c_{a d} \Gamma^d_{b c} \ .
\ee
A useful formula is that for the metric ansatz 
\be
G_{ab} =   \left( \begin{array}{c    c}  \tilde{h}_{ij} & 0 \\ 0 & e^{2 \g} 
  \end{array} \right) \ ,
\ee
the Ricci scalar is given by
\be
R[G] = R[\tilde{h} ] - 2 (\pd \g)^2 -2 \tilde{\nabla}^2 \g  \  .
\ee

Under a conformal rescaling $ \tilde{h}_{ij} = e^{2 \om} h_{ij}$ we have the following  useful identities
\ba
\label{Weylresc}
\tilde{\nabla}^2 \Phi &= & e^{-2 \om} \left(  \nabla^2 \Phi  + (d-2) \pd_i \om \pd^i \Phi   \right) \ , \\
R[\tilde{h}] &=&  e^{-2 \om}\left( R[h] - 2(d-1) \nabla^2 \om - (d-1)(d-2) \pd_i \om \pd^i \om  \right)     \  .
\ea

\end{appendix}


\providecommand{\href}[2]{#2}\begingroup\raggedright\endgroup

\end{document}